\documentclass[%
 reprint,
 amsmath,amssymb,
 aps,pra
]{revtex4-2}
\usepackage{graphicx}
\usepackage{amsmath}
\usepackage{nicefrac}
\usepackage[hidelinks]{hyperref}
\usepackage{braket}
\usepackage{amsmath}
\usepackage{siunitx}

\begin{document}
\title{
Splitting and connecting singlets in atomic quantum circuits
}
\author{Zijie Zhu}
\author{Yann Kiefer}
\author{Samuel Jele}
\author{Marius Gächter}
\author{Giacomo Bisson}
\author{Konrad Viebahn}
\email[]{viebahnk@phys.ethz.ch}
\author{Tilman Esslinger}
\affiliation{Institute for Quantum Electronics \& Quantum Center, ETH Zurich, 8093 Zurich, Switzerland}

\begin{abstract}
Gate operations composed in quantum circuits form the basis for digital quantum simulation and quantum processing.
While two-qubit gates generally operate on nearest neighbours, many circuits require nonlocal connectivity and necessitate some form of quantum information transport.
Yet, connecting distant nodes of a quantum processor still remains challenging, particularly for neutral atoms in optical lattices.
Here, we create singlet pairs of two magnetic states of fermionic potassium-40 atoms in an optical lattice and use a bi-directional topological Thouless pump to transport, coherently split, and separate the pairs, as well as to demonstrate interaction between them via tuneable {\sc $($swap$)^\alpha$}-gate operations. 
We achieve pumping with a single-shift fidelity of 99.78(3)\% over 50 lattice sites and split the pairs within a decoherence-free subspace. Gates are implemented by superexchange interaction, allowing us to produce interwoven atomic singlets. For read-out, we apply a magnetic field gradient, resulting in single- and multi-frequency singlet-triplet oscillations. Our work shows avenues to create complex patterns of entanglement and new approaches to quantum processing, sensing, and atom interferometry.
\end{abstract}

\maketitle

The connectivity of gate operations is central in quantum information processing. Optical lattices, in which a large number of atoms can be prepared and positioned in well-defined quantum states~\cite{ludlow_optical_2015,daley_practical_2022,hartke_quantum_2022}, including control over tunnelling, have a unique potential for scaling up quantum processors. 
However, connecting spatially separated quantum states on this platform has remained a challenge~\cite{jaksch_entanglement_1999,daley_quantum_2008}.
Lattice-based atom shuttling, either via state-dependent light shifts~\cite{mandel_controlled_2003,steffen_digital_2012,robens_low-entropy_2017,kumar_sorting_2018}, or via additional scanning tweezers~\cite{young_atomic_2024,shaw_erasure_2025,norcia_iterative_2024,gyger_continuous_2024}, suffer from motional heating or atom loss.
In this article, we introduce bidirectional topological Thouless pumping~\cite{oka_floquet_2019,citro_thouless_2023,nakajima_topological_2016,lohse_thouless_2016,koepsell_robust_2020,minguzzi_topological_2022,hu_topological_2020} in combination with controlled superexchange interactions~\cite{trotzky_time-resolved_2008,greif_short-range_2013,dai_generation_2016,zhang_scalable_2023,barmettler_quantum_2008,vaucher_creation_2008} as a coherent toolbox to transport, separate, interweave, and recombine atomic quantum states in an optical lattice.

\begin{figure*}[t!]
    \includegraphics[width=1\textwidth]{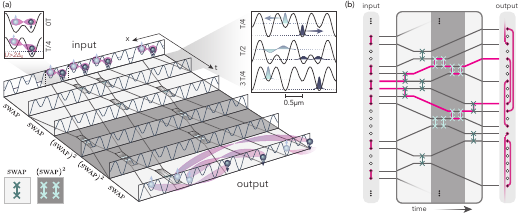}
         \caption{\textbf{Quantum circuits based on topological pumping for splitting and connecting atomic singlet pairs in an optical superlattice.} (a) Schematic of the experimental architecture, showing an exemplary circuit of depth five consisting of several parallel ${\text{\sc swap}}$ and $({\text{\sc swap}})^2$ gates.
         Left inset: input state preparation of an atomic singlet $\ket{s}$ occupying two separate bands.
         Right inset: illustration of one operation cycle of the Thouless pump.
         Atoms in the ground (right-moving, Chern number $C = +1$) and first excited (left-moving, $C = -1$) bands each feature quantised and state-independent displacement in opposite directions.
         As a result, singlet pairs are separated by many lattice sites.
         (b) Circuit representation of the process displayed in panel (a) whose input consists of randomly positioned singlets in the lattice (filled dots and magenta lines).
        }
    \label{fig:1}
\end{figure*}

We create entangled Bell pairs by splitting and separating atomic spin singlets, which are abundant in the quantum-degenerate gas of fermionic potassium atoms in the optical lattice.
The dynamical lattice potential exhibits two distinct energy bands, whose orthogonal orbitals (Wannier states) shift in opposite directions upon adiabatic modulation, forming a state-independent topological Thouless pump with Chern numbers $C = \pm 1$~\cite{walter_quantization_2023,zhu_reversal_2024,viebahn_interactions_2024} [Fig.~\ref{fig:1}(a)].
We prepare the pairs so that they occupy both orbitals, therefore, their two components are spatially separated during pumping.

Now consider two pairs that are initially located several lattice sites apart.
While the pairs' centers of mass remain at the same distant position, each pumping cycle increases the separation of the two components of each pair.
In our one-dimensional geometry, the inward-facing components of the pairs are pumped closer together until they collide when they share the same unit cell.
In this manner, we are able to connect distant quantum states in the lattice.
Furthermore, the precise control over superexchange interactions enables programmable {\sc $($swap$)^\alpha$}-gate operations between two atoms, allowing us to move quantum states through each other using {\sc swap} gates and, more generally, paving the way for the generation of complex entanglement patterns [Fig.~\ref{fig:1}(b)].

\begin{figure*}[t!]
    \includegraphics[width=1\textwidth]{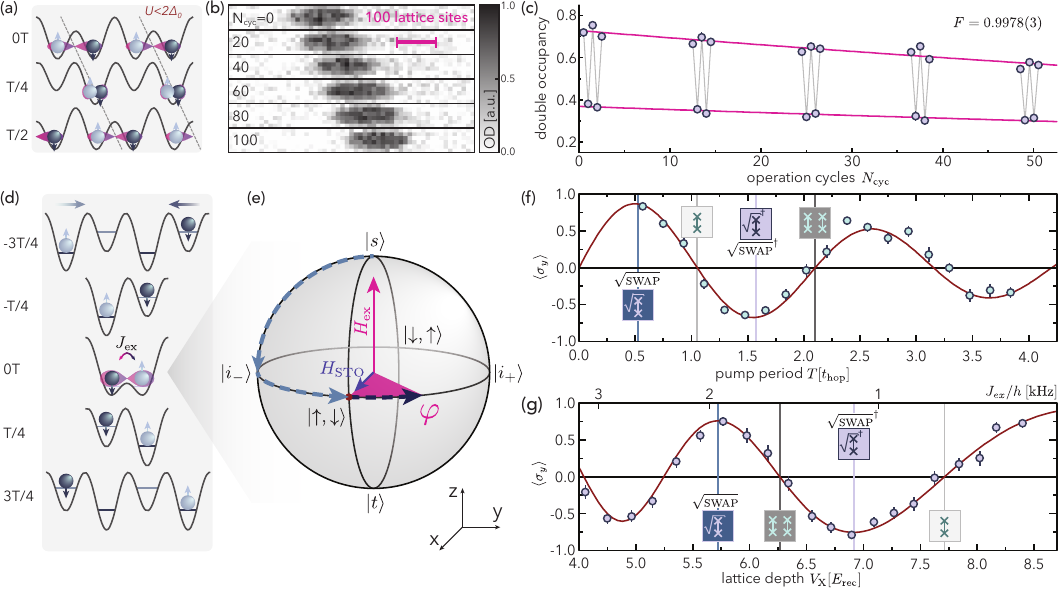}
         \caption{\textbf{Shuttle and gate operations enabled by topological pumping and controlled superexchange interactions}. (a) Illustration of shuttle operations for noninteracting atomic singlets via topological pumping. 
         (b) In-situ absorption images of the atomic cloud during the shuttle operation for different numbers of operation cycles $N_{\text{cyc}}$.
         (c) Measured double occupancy 
        as function of $N_{\text{cyc}}$, including one pump reversal halfway.
        The low and high double occupancies correspond to the dimerised ($\Delta=0$) and staggered configurations ($\Delta=\Delta_0$), respectively.
         The solid lines denote the respective exponential fits of the data in the two configurations, yielding an operation fidelity of $F=0.9978(3)$. 
         (d) Visualisation of the configurable gate operations achieved by controlling the superexchange interaction $J_{\text{ex}}$ when two atoms meet during pumping. 
         The example shows a \textsc{swap} gate.
         (e) Two-particle Bloch sphere displaying the Hamiltonians $H_{\text{ex}}$ and $H_{\text{STO}}$. The remaining equatorial states are $\ket{i_{-}}=\left(\ket{\downarrow,\uparrow}-i\ket{\uparrow,\downarrow}\right)/\sqrt{2}$ and $\ket{i_{+}}=\left(\ket{\downarrow,\uparrow}+i\ket{\uparrow,\downarrow}\right)/\sqrt{2}$. The initialisation of the product state $\ket{\uparrow,\downarrow}$ is indicated by light dashed lines.
         The time evolution of the state under the superexchange Hamiltonian $H_{\text{ex}}$ corresponds to a rotation around the $z$-axis by an angle $\varphi$ (dashed purple line). 
         Gate operations are controlled by adjusting $\varphi$, which is determined by measuring the projection of the final state onto the $y$-axis of the Bloch sphere (Appendix). 
         We realise the gates as a function of the pump period \( T \) in units of the average tunnelling over one period, \( t_{\text{hop}} \) [(f)], and as a function of the optical lattice depth \( V_{\text{X}} \) [(g)].
         The upper axis in (g) corresponds to the strength of $J_{\text{ex}}$ in the dimerised configuration. The insets indicate the parameters used for the implementation of the respective gates.
         The error bars in (c), (f), and (g) correspond to the standard error for at least 9, 14, and 12 repetitions, respectively.
         The lines in (f) and (g) are damped sinusoidal fits with respect to $T$ ($J_{\text{ex}}$).}
    \label{fig:2}
\end{figure*}

The atomic spin space is spanned by two magnetic sublevels of potassium-40 within the $F = 9/2$ manifold and we denote $m_F = -9/2$ as $\downarrow$ and $m_F = -5/2$ as $\uparrow$.
The entangled Bell pairs $\ket{s}=\left(\ket{\downarrow,\uparrow}-\ket{\uparrow,\downarrow}\right)/\sqrt{2}$ (atomic `Heisenberg' singlet, anti-symmetric) and $\ket{t}=\left(\ket{\downarrow,\uparrow}+\ket{\uparrow,\downarrow}\right)/\sqrt{2}$ (triplet, symmetric) form a decoherence-free subspace, which is insensitive to magnetic noise and motional perturbations~\cite{kielpinski_decoherence-free_2001,anderlini_controlled_2007,kaufman_entangling_2015}.
We define $\ket{\uparrow,\downarrow}$ as a state where \mbox{spin-$\uparrow$} and spin-$\downarrow$ each occupy distinct orbitals, such as the left and right sites of a double-well.
The singlet state $\ket{s}$ remains invariant under SU(2) transformations and it is thus independent of spin convention (up to an irrelevant global phase).
The states $\ket{s}$ and $\ket{t}$ may be coupled via a magnetic gradient, leading to singlet-triplet oscillation (STO) which we use for read-out.
In particular, the energy splitting between $\ket{\downarrow,\uparrow}$ and $\ket{\uparrow,\downarrow}$ caused by the magnetic gradient increases with spatial separation of the two magnetic moments and the oscillation frequency between the singlet and the triplet states increases accordingly~\cite{trotzky_controlling_2010,greif_short-range_2013,taie_observation_2022}.

The atoms are tightly confined in a three-dimensional dynamic lattice potential which creates an array of independent one-dimensional tubes in the $x$-direction [Fig.~\ref{fig:1}(a)].
The pumping lattice is characterized by tunnelling parameters $t_{x}$ and $t_{x}'$ on alternating bonds, and alternating lattice sites exhibit an offset energy $\pm\Delta$.
Thouless pumping is achieved via a periodic modulation of the lattice potential, cycling through the staggered ($\Delta=\Delta_{0}$, $t_{x}=t_{x}'$) and the dimerised ($\Delta=0$, $t_{x}\neq t_{x}'$) configurations, which adiabatically shuttles the atoms within the lattice (Appendix).
The interparticle interaction strength is characterised by the on-site Hubbard $U$.
Strong attractive interactions during lattice loading ensure that between 60\% and 75\% of the total number of $5.3(2)\times10^{4}$ atoms start off in doubly occupied unit cells.

In a first experiment, we assess the shuttling fidelity by transporting atomic singlet pairs as a whole (rather than splitting them) which occurs as long as the Hubbard $U$ remains smaller than a critical value of $2\Delta_{0}$~\cite{walter_quantization_2023}.
The resulting singlet pairs exhibit the same anti-symmetry in spin space as the Heisenberg singlets $\ket{s}$, yet, due to the lack of strong fermion repulsion both atoms occupy the same orbital.
These singlets can be shuttled together by one lattice site every half pump period $T$, which we define as one operation cycle, confirmed by in-situ measurements of the cloud displacement (Fig.~\ref{fig:2}(b), $U = 0$).
Unless otherwise specified, we used a fixed cycle time of $T/2 = \SI{1.25}{ms}$ in this work.
This gate speed is not fundamental, but rather limited by technical constraints of our phase lock bandwidth~\cite{walter_quantization_2023}.
During the shuttling process, double occupancy, i.e., two atoms occupying the same lattice site, varies in time since the ground state orbital becomes alternatingly distributed over two sites and localised on one site [Fig.~\ref{fig:2}(a)].
In particular, the double occupancy reaches its maximum in the staggered lattice configuration ($\Delta=\pm\Delta_0$) and its minimum in the dimerised configuration ($\Delta=0$), as shown in Fig.~\ref{fig:2}(c) ($U = 0$).
Since our spectroscopic measurement of double occupancy is ground-state-selective (Appendix), this provides a microscopic observable to evaluate the fidelity of the shuttling operation.
A decrease in double occupancy as function of the number of operation cycles $N_{\text{cyc}}$ indicates that an atom from a singlet pair has spread to neighbouring sites, has been excited to higher bands, or otherwise been lost.
The experimental sequence for Fig.~\ref{fig:2}(c) involves pumping forward and then returning to the original position by reversing the pump direction, ensuring consistent detection efficiency in the centre of the trap.
Exponential fits of the double occupancy in the staggered and dimerised lattice configurations yield a fidelity of $F=0.9978(3)$ for a single-atom shuttle operation by one lattice site (Appendix).
In all subsequent experiments we ramp the interactions to the strongly repulsive Heisenberg regime during loading ($U\gg t_x, t_x',\Delta$) in order to suppress double occupancies and adiabatically transfer the two constituents of each singlet pair to separate orbitals (Fig.~\ref{fig:1}(a), left inset).

\begin{figure*}[t!]
    \includegraphics[width=1\textwidth]{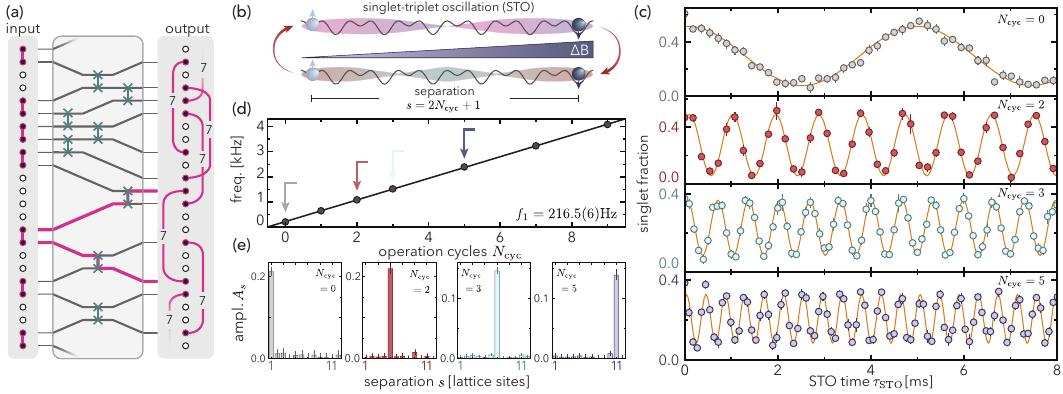}
         \caption{\textbf{Determining the separation between two entangled atoms with singlet-triplet oscillations (STOs).} 
         (a) All-{\sc swap} circuit of depth three ($N_{\text{cyc}}=3$), resulting in a separation of $s=7$ lattice sites.
         The magenta trace within the circuit illustrates the path of an exemplary atomic singlet.
         (b) Schematic of singlet-triplet oscillations of atoms split by multiple lattice sites. A magnetic gradient $\Delta B$ lifts the degeneracy between $\ket{\downarrow,\uparrow}$ and $\ket{\uparrow,\downarrow}$ states, leading to an oscillation between the singlet and triplet states. The oscillation frequency is proportional to the separation $s=2N_{\text{cyc}}+1$. (c) Measurement of the STO frequency for different numbers of operation cycles $N_{\text{cyc}}=0,2,3,$ and $5$ (from top to bottom) with {\sc swap} gates, corresponding to separations up to $11$ sites. 
         The orange lines denote sinusoidal fits to the data and the error bars are the standard error for 3-6 repetitions. 
         (d) STO frequencies as a function of the number of operation cycles $N_{\text{cyc}}$. The coloured arrows point to the respective time traces in (c). Time traces for $N_{\text{cyc}}=1,7,$ and $9$ are plotted in Fig.~\ref{fig:S6}. The solid black line is a proportional fit to the data, used to determine the STO base frequency $f_{1}$.
         The error bars, representing the uncertainty of the sinusoidal fit, are smaller than the data points.
         (e) Distribution of frequency components, corresponding to different values of atom separation $s$. These distributions are evaluated by applying a twelve-frequency sinusoidal fit to the four time traces in (c) with frequencies $s \times f_{1}$ up to $s=12$. The amplitudes $A_{s}$ of the respective components are plotted as function of $s$, where error bars denote the amplitude uncertainty of the fit.}
    \label{fig:3}
\end{figure*}

Interparticle interactions allow the programmable application of gate operations when atoms move in opposite directions and meet, instead of all moving in the same direction.
As illustrated in Fig.~\ref{fig:2}(d), such atoms colliding on one unit cell interact via superexchange whose strength is characterised by $J_{\text{ex}} = 4t_x^2/\left[U\left(1-(2\Delta/U)^2\right)\right]$~\cite{duan_controlling_2003} (Appendix).
Depending on the lattice depths used in this work, the intra-dimer tunnelling in the dimerised configuration is betweeen \SI{1}{kHz} and \SI{2}{kHz}.
The maximum site offset $\Delta$ reaches \SI{2}{kHz}, and the Hubbard $U$ is approximately \SI{8}{kHz} (Fig.~\ref{fig:S1}).
The value of $J_\mathrm{ex}(\tau)$ varies by orders of magnitude during pumping and it vanishes periodically, giving rise to discrete gate operations for each operation cycle. Meanwhile, superexchange interactions between atoms in different double wells remain negligible.
Superexchange gate operations can be represented on a two-particle Bloch sphere [Fig.~\ref{fig:2}(e)].
The fermionic superexchange Hamiltonian $\hat{H}_{\text{ex}}=J_{\text{ex}}\hat{\sigma}_{z}/2$ induces a rotation around the $z$-axis by an angle $\varphi=\frac{1}{\hbar}\int_{-T/4}^{T/4}J_{\text{ex}}(\tau)\,d\tau$, where $\hat{\sigma}_{z}$ is the third Pauli matrix.
Precise control of $\varphi$ enables the realisation of the entire family of partial swap gates, such as $\sqrt{\text{\sc swap}}$ ($\varphi=\pi/2$), \textsc{swap} ($\varphi=\pi$), $\sqrt{\text{\sc swap}}^{\dag}$ ($\varphi=3\pi/2$), and $({\text{\sc swap}})^2$ ($\varphi=2\pi$), see also refs.~\cite{anderlini_controlled_2007,zhang_scalable_2023}.
Here we have simplified the gate action by considering only two opposite spins.
However, since all three triplet states are degenerate, the realisation of the partial swap gates remains valid in the full two-particle Hilbert space, including $\ket{\downarrow,\downarrow}$ and $\ket{\uparrow,\uparrow}$ (Appendix).
The entangling gate $\sqrt{\text{\sc swap}}$, combined with single-qubit rotations~\cite{weitenberg_single-spin_2011}, allows the construction of a universal gate set for quantum computation~\cite{loss_quantum_1998}.

The indistinguishable nature of the fermionic constituents enables two complementary (but equivalent) interpretations of the action of two-particle gates.
In the first interpretation, operations act on the spin-$\{\uparrow,\downarrow\}$ degrees of freedom and a {\sc swap} gate can be understood as two atoms swapping their spin states.
Strongly repulsive interactions prevent atoms from passing through each other, allowing individual atoms to be labelled in positional order, which could be used to construct a quantum circuit representation based on effectively distinguishable qubits.
In the second, or `motional', interpretation a {\sc swap} operation exchanges the atoms' positions and atoms can become delocalised in the lattice through actions such as $\sqrt{\text{\sc swap}}$, which is relevant for fermionic quantum computing~\cite{bravyi_fermionic_2002}.
In contrast to atom shuttling in optical tweezers~\cite{beugnon_two-dimensional_2007,graham_multi-qubit_2022,bluvstein_logical_2023} or ion traps~\cite{moses_race-track_2023}, topological pumping maintains motional ground-state coherence during transport.

In order to calibrate the gate operations and determine the rotation angle $\varphi$, we first prepare atom pairs in the product state $\ket{\uparrow,\downarrow}$. Next, we pump for one half period, realising one gate operation as shown in Fig.~\ref{fig:2}(d) ($-T/4$ to $T/4$), during which the state evolves on the equator of the two-particle Bloch sphere. Finally, we determine $\varphi$ by measuring the projection of the resulting state along the $y$-axis of the Bloch sphere (Appendix).
We present two different methods to engineer the gate operations. In Fig.~\ref{fig:2}(f), we vary the pump period $T$ to control the interaction duration. 
In Fig.~\ref{fig:2}(g), we realise different gates by adjusting the lattice depth $V_{\text{X}}$ and thus the superexchange coupling $J_{\text{ex}}$. 
The insets indicate the parameters used for implementing the respective gates, determined from the fitted curves shown in solid red lines.

\begin{figure*}[t!]
    \includegraphics[width=1\textwidth]{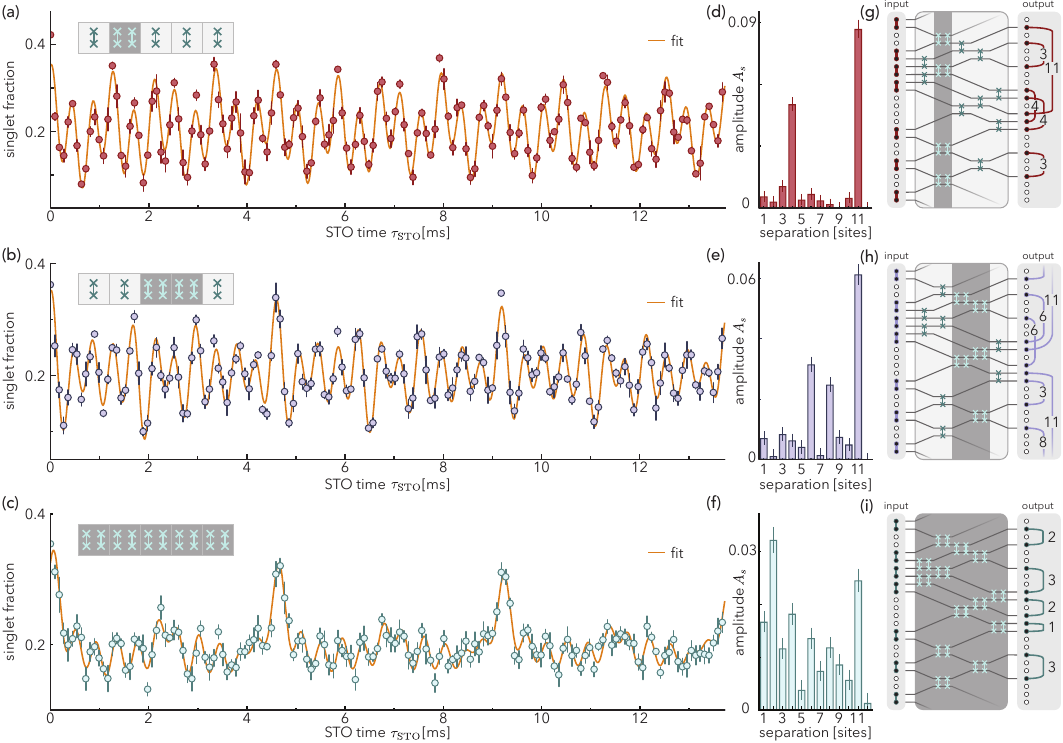}
         \caption{\textbf{Quantum states generated by configurable gate sequences exhibiting multi-frequency singlet-triplet oscillations.} 
         (a-c) Singlet fraction as a function of STO time $(\tau_{\text{STO}})$ after a total of five operation cycles, each containing one [(a)], two [(b)], or all [(c)] ({\sc swap})$^2$ gates, while all remaining gates are {\sc swap} gates.
         The order of the respective gates is visualised in the insets. The orange lines represent twelve-frequency sinusoidal fits of the data with a base frequency of $f'_{1}=\SI{218(1)}{Hz}$. The error bars denote the standard error for at least 3 [(a, b)] or 6 [(c)] experimental repetitions, respectively. 
         (d-f), Amplitudes $A_s$ of the corresponding fits, directly revealing atomic singlets separated by different distances. The error bars correspond to the uncertainty of the fit. (g-i), Circuit representations of the gate sequences implemented in (a-c).
         }
    \label{fig:4}
\end{figure*}

With full control over shuttle and gate operations, we are able to compile quantum circuits with the aim of splitting and connecting singlet pairs within the lattice.
The first circuit we realise consists solely of \textsc{swap} gates, which can be envisioned as two atoms exchanging positions or passing through each other, which is useful for controlling connectivity.
As a result, the output state consists of interwoven singlet pairs, each separated by $s=2N_{\text{cyc}}+1$ lattice sites, where $N_{\text{cyc}}$ is the circuit depth (i.e.~the number of operation cycles).
In Fig.~\ref{fig:3}(a), we show an exemplary circuit of depth $N_{\text{cyc}}=3$, outputting singlet pairs separated by 7 sites.
To measure the separation between two entangled atoms from an initial singlet pair, we employ singlet-triplet oscillations (STOs).
In short, a magnetic gradient $\Delta B$ is applied for a specified time $\tau_{\text{STO}}$ which causes an energy offset $\Delta_{\uparrow\downarrow} \propto \Delta B \times s$ between the states $\ket{\downarrow,\uparrow}$ and $\ket{\uparrow,\downarrow}$, corresponding to the Hamiltonian $\hat{H}_{\text{STO}}=\Delta _{\uparrow\downarrow}\hat{\sigma}_{x}/2$ on the two-particle Bloch sphere [Fig.~\ref{fig:2}(e)], where $\hat{\sigma}_{x}$ is the first Pauli matrix. 
The oscillation frequency $\Delta_{\uparrow\downarrow}/h$ between singlet and triplet states thus reveals the atom separation $s$ [Fig.~\ref{fig:3}(b)]. 
After the STO, we reverse the pump to bring the entangled pairs back together for detection.
This sequence can be viewed as a many-particle interferometer.

In Fig.~\ref{fig:3}(c), we present STOs for increasing circuit depths up to $N_{\text{cyc}}=5$, corresponding to increasing atom separations $s$. We determine the oscillation frequency by fitting a sinusoidal curve to each time trace.
The results are shown in Fig.~\ref{fig:3}(d), along with additional data up to $s=19$ sites, whose corresponding time traces are plotted in Fig.~\ref{fig:S5}.
All measurements in Fig.~\ref{fig:3}(c) and Fig.~\ref{fig:S6} are performed under the same gradient $\Delta B$, whereas a reduction in contrast as function of $s$ is attributed to fluctuating lattice depths and resulting deviations from perfect \textsc{swap} operations.
A proportional fit of the data yields a slope of $f_{1}=\SI{216.5(6)}{Hz}$, which gives the STO base frequency corresponding to entangled pairs on adjacent lattice sites. Pairs separated by $s$ sites thus exhibit STOs at a frequency of $s\times f_{1}$. We then map the time trace to the distribution of separations $s$ by calculating the Fourier spectrum of the time trace.
Given that $s$ must be an integer in lattice systems, we apply a multi-frequency sinusoidal fit $F_{\text{singlet}}(\tau)=\Sigma_{s}A_{s}{\text{sin}}(2\pi s f_{1}\tau+\theta_{s})$ with a global damping and offset.
The amplitude $A_{s}$ is then proportional to the fraction of singlet pairs separated by a certain distance $s$.
The dominating peaks in Fig.~\ref{fig:3}(e), shown up to 11 lattice sites ($N_{\text{cyc}} = 5$), demonstrate the ability to coherently split fermionic singlet pairs by programmable distances.

We now turn to more complex quantum circuits to connect and recombine many different singlets at various distances via combinations of \textsc{swap} and $({\text{\sc swap}})^2$ gates.
A $({\text{\sc swap}})^2$ gate does not alter the spin states when two atoms meet and can thus be considered a reflection, leading to intricate STO waveforms and a non-trivial distribution of singlet pairs.
In Fig.~\ref{fig:4}, we present the measured STOs of three different circuits with constant circuit depth ($N_{\text{cyc}}=5$).
In the first example, we insert a single parallel $({\text{\sc swap}})^2$ operation at the second position of an otherwise purely \textsc{swap} sequence, leading to the emergence of multiple frequency components in the STO signal [Fig.~\ref{fig:4}(a)]. We analyse the time trace analogously to Fig.~\ref{fig:3}(e), using an independently calibrated base frequency of $f_{1}'=\SI{218(1)}{Hz}$.
The resulting amplitudes $A_{s}$, representing the proportion of singlet pairs separated by $s$ lattice sites, are shown in Fig.~\ref{fig:4}(d).
We also apply a fast Fourier transform to the time trace (Fig.~\ref{fig:S6}), which agrees with the fit.
A significant contribution at $s=4$, in addition to $s=2\times5+1=11$ lattice sites, can be understood as follows.
During the initial operation cycle, two atoms originating from a singlet pair are separated by $2\times1+1=3$ lattice sites while in the second operation cycle, one atom can be reflected while the other continues moving, resulting in a total separation of $3+1=4$ sites.
From then on, the two atoms shift in the same direction and their separation remains constant.
If both atoms originating from a singlet pair are reflected in the second operation cycle, they will be shuttled in opposite directions during the next three cycles, resulting in a separation of $\left| 3-2\times3 \right|=3$ lattice sites.
This scenario is considerably less probable, in agreement with the low measured value at $s=3$. 
Both scenarios can be identified with trajectories in the circuit diagram shown in Fig.~\ref{fig:4}(g).
The occurrence count of specific distances in the schematic (output) does not directly reflect the amplitudes $A_{s}$, as it depicts only an examplary subset of the entire system.
While STOs have been also observed on neighbouring lattice sites with SU(N) fermions~\cite{taie_observation_2022}, in our experiment the occurrence of multiple frequencies is a direct consequence of the distribution of spatially separated singlet pairs.

Changing the gate composition of quantum circuits leads to additional contributions to the multi-frequency STOs.
For instance, Fig.~\ref{fig:4}(b) shows the time trace for a situation with two parallel $({\text{\sc swap}})^2$ gates, at the third and fourth position in the sequence.
The process involving a single reflection at $N_{\text{cyc}}=3$ or 4 results in two major contributions to the STO signal, corresponding to $s=6$ and $s=8$ lattice sites.
The process involving multiple reflections gives rise to signals at $s=1, 3, 4$, etc.
In addition, we implement a quantum circuit consisting solely of $({\text{\sc swap}})^2$ gates.
In this configuration, neighbouring singlets are constraining each other, generally resulting in small separations.
The time trace [Fig.~\ref{fig:4}(c)] and the fitted amplitudes $A_{s}$ [Fig.~\ref{fig:4}(f)] confirm that the multi-frequency STO is dominated by low-frequency components, corresponding to small values of $s$.
The remaining fraction of singlets separated by $s=11$ lattice sites can be attributed to low-density regions near the edges of the system and imperfections in the gate operations.
The striking revival feature in the all-$({\text{\sc swap}})^2$ time trace [Fig.~\ref{fig:4}(c)] shows the harmonicity of the frequencies and their phase coherence.
Considering that the signal is contributed by more than ten thousand individual entangled pairs, such phase coherence highlights the potential of topological pumping for scaling up quantum circuits.

In conclusion, we have experimentally achieved high-fidelity shuttle operations of fermionic singlets using topological pumping in an optical lattice.
Both motional ground-state coherence and the entanglement in the spin sector remain preserved during pumping, enabling us to transport entangled atoms together or separate them by tens of lattice sites. Utilising superexchange interactions, we further realise two-qubit {\sc $($swap$)^\alpha$} gates, facilitating programmable quantum circuits and connecting distant singlet pairs. 
Our platform is compatible with single-atom control techniques such as optical tweezers, enabling flexible operations including local spin flips and orbital inversion via gapless edge modes~\cite{zhu_reversal_2024}. These capabilities allow reconfigurable atom trajectories and support more complex, adaptable gate sequences.
Compared to other platforms targeting universal quantum computation, neutral atoms in lattices have a natural advantage in terms of power utilisation, atomic qubit density, scalability, and the ability to perform parallel operations.
By improving the connectivity in optical lattices using topological pumping, our work opens up new possibilities for quantum information processing based on atoms and molecules.
This includes applications in fermionic quantum computing~\cite{bravyi_fermionic_2002,gonzalez-cuadra_fermionic_2023}, symmetry-protected operations~\cite{freedman_symmetry_2021,rudolph_two-qubit_2023}, singlet-triplet qubits~\cite{petta_coherent_2005}, as well as computing based on quantum walks~\cite{young_atomic_2024,childs_universal_2013}.\\

\section*{Acknowledgements}

We would like to thank Alex Baumgärtner and Peter Zoller for comments on a previous version of the  manuscript.
We thank Alexander Frank for assistance with electronics equipment.
We acknowledge funding by the Swiss National Science Foundation (Grant No.~200020\_212168, Advanced grant TMAG-2\_209376, 20QT-1\_205584, as well as Holograph UeM019-5.1).


%

\clearpage

\setcounter{figure}{0} 
\setcounter{equation}{0}

\renewcommand\thefigure{A\arabic{figure}} 
\renewcommand\thetable{A\arabic{table}}

\section*{Appendix}

\subsection*{Superlattice potential}

The time-dependent optical lattice potential is given by
\begin{equation}
\begin{split}\label{eq:potential}
    V(&x,y,z,\tau) = \\
    &-V_\mathrm{X}\cos^2(kx+\theta/2)\\
    &-V_\mathrm{Xint}\cos^2(kx)\\
    &-V_\mathrm{Y}\cos^2(ky)\\
    &-V_\mathrm{Z}\cos^2(kz)\\
    &-\sqrt{V_\mathrm{Xint}V_\mathrm{Z}}\cos(kz)\cos(kx+\varphi_{\text{SL}}(\tau))\\
    &-I_\mathrm{XZ}\sqrt{V_\mathrm{Xint}V_\mathrm{Z}}\cos(kz)\cos(kx-\varphi_{\text{SL}}(\tau)),
\end{split}
\end{equation}
where $k=2\pi/\lambda$, $\lambda=\SI{1064}{\nano\meter}$ and the imbalance factor is given by $I_\mathrm{XZ} = 0.777(3)$. The lattice depths $V_\mathrm{X}$, $V_\mathrm{Xint}$, $V_\mathrm{Y}$ and $V_\mathrm{Z}$ are listed in Table \ref{tab:lattice_depths} in units of the recoil energy $E_{\mathrm{rec}} = h^2/2m\lambda^2$, where $m$ is the atomic mass.
In our experiments, the lattice depth along the $y$- and $z$-directions is sufficiently large to effectively freeze the dynamics in these two directions.
The resulting potential creates a superlattice structure along the $x$-direction, characterised by staggered tunnellings $t_{x}(\tau)$ and $t_{x}'(\tau)$, as well as a staggered site offset $\Delta(\tau)$ (Fig.~\ref{fig:S1}a).
A simple linear ramp of the superlattice phase $\varphi_{\text{SL}}$, imprinted via an acousto-optic modulator, leads to a periodic modulation of the model parameters with a period $T$ (Fig.~\ref{fig:S1}b).

{\renewcommand{\arraystretch}{1.2}
\begin{table*}[ht]
    \centering
    \begin{tabular}{rclll}
        \hline\hline
         & $V_\mathrm{X}$[$E_\mathrm{rec}$] & $V_\mathrm{Xint}$[$E_\mathrm{rec}$] & $V_\mathrm{Y}$[$E_\mathrm{rec}$] & $V_\mathrm{Z}$[$E_\mathrm{rec}$]\\
        \hline
        Fig.~\ref{fig:2}(b),c & 7.539(5) & 0.256(1) & 32.35(2) &29.42(3)\\
        Fig.~\ref{fig:2}(f) & 7.484(2) & 0.2521(7) & 27.047(2) & 27.191(3)\\
        Fig.~\ref{fig:2}(g) & 4.555(8),\dots,8.40(1) & 0.2500(4) & 26.988(2) & 27.50(2)\\
        Fig.~\ref{fig:3}, Fig.~\ref{fig:S2} & 7.603(5) & 0.261(2) & 27.02(2) & 27.03(2)\\
        Fig.~\ref{fig:4} & 6.012(5), 7.484(9) & 0.249(2) & 27.05(1) & 27.20(1)\\
        \hline
    \end{tabular}
    \caption{Lattice depths used in the experiment. In Fig. \ref{fig:2}(g), $V_{\text{X}}$ was scanned over the range indicated in the table. For Fig.~\ref{fig:4}, the two values correspond to the lattice depth $V_{\text{X}}$ used to realise the $({\text{\sc swap}})^2$ and ${\text{\sc swap}}$ gates. The values in the brackets denote the respective standard errors.}
    \label{tab:lattice_depths}
\end{table*}
}

\subsection*{Topological pumping}
The periodic modulation of the lattice potential can be considered as a static `short' lattice with a moving `long' lattice, describing a Thouless pump~\cite{citro_thouless_2023} (Fig.~\ref{fig:S1}a).
Compared to its classical counterpart, the quantum nature of the Thouless pump is manifested in its directional dependence of the motional states.
In a band-structure picture, the pump in a bipartite one-dimensional lattice potential features two topologically distinct bands.
The topological properties of these bands are characterised by Chern numbers. These numbers are derived by mapping the space- and time-periodic Hamiltonian onto a time-independent 2D Harper-Hofstadter-Hatsugai (HHH) model, which incorporates both a real and a synthetic dimension.
The two lowest bands of the HHH model have Chern numbers $C = \pm 1$ giving rise to quantised transport of two lattice sites per period in opposite directions~\cite{oka_floquet_2019,zhu_reversal_2024}.

In contrast to bichromatic setups of topological pumps~\cite{nakajima_topological_2016,lohse_thouless_2016,koepsell_robust_2020}, the depth of the `moving' lattice in our case is also periodically modulated. This modulation occurs automatically due to the time dependence of the two terms proportional to $\sqrt{V_\mathrm{Xint}V_\mathrm{Z}}$ in Eq.~\ref{eq:potential} and requires no change in laser intensities.
The modulation ensures a smoother time evolution of the topological bandgap and it increases the duration of the superexchange interaction (width of the feature in Fig.~\ref{fig:S1}c).


\subsection*{Realisation of {\sc $($swap$)^\alpha$} gates}

\begin{figure}[htbp]
    \includegraphics[width=0.48\textwidth]{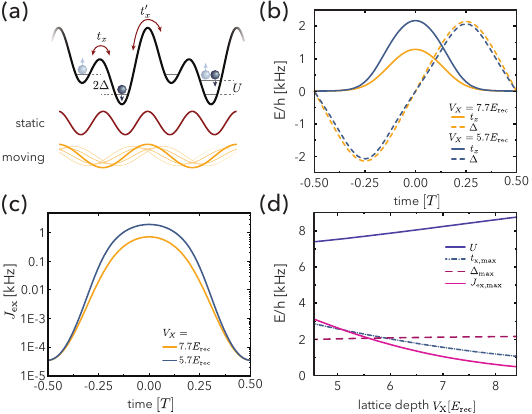}
         \caption{\textbf{Optical lattice potential and resulting Hamiltonian parameters.} (a) 
         The superposition of a short, static (shown as red) and a long, moving (shown as yellow) lattice creates the bipartite superlattice potential (black) with tunnellings $t_x$, $t_x'$, and site-offset $\Delta$. The lattice is formed of a single laser wavelength $\lambda = \SI{1064}{nm}$ and the colours merely illustrate the two main contributions to the pumping lattice.
         Double occupancies are suppressed due to a large repulsive on-site energy $U>2\Delta_0$.
         (b) Dependence of tunnelling $t_x$ and site-offset $\Delta$ variations within one pump cycle for two different lattice depths $V_\mathrm{X}=\{5.7, 7.7\} E_{\text{rec}}$.
         The tunnelling $t_x'$ is omitted for simplicity, as $t_x' = t_x(\tau + T/2)$. 
         (c) Calculated superexchange energy $J_\mathrm{ex}$ as a function of time within one pump cycle for two different lattice depths $V_{\text{X}}$. The superexchange energy $J_\mathrm{ex}$ is negligible in the staggered lattice configuration ($\pm 0.25T$) due to large $\Delta$ and small $t_x$. It takes on larger values when entering and exiting the balanced double-well configuration ($\Delta=0$). (d) Maximum tunnelling $t_x$, site-offset energy $\Delta$, interaction strength $U$, and resulting maximum of $J_\mathrm{ex}$ as a function of lattice depth $V_{\mathrm{X}}$. 
         }
    \label{fig:S1}
\end{figure}

We model two particles meeting in a unit cell by considering a double-well with sites labelled by L and R. With Hubbard interactions $U$, the corresponding Hamiltonian is given by
\begin{equation}
\begin{split}
    \hat{H}_\mathrm{DW} = -t_x&\sum_{\sigma=\uparrow,\downarrow}(\hat{c}^\dagger_{\mathrm{L}\sigma}\hat{c}_{\mathrm{R}\sigma}+\mathrm{h.c.})\\ 
    +&\Delta\sum_{\sigma=\uparrow,\downarrow} \left(\hat{n}_{\mathrm{L}\sigma} -\hat{n}_{\mathrm{R}\sigma}\right)
    + \sum_{\alpha=\mathrm{L,R}} U\hat{n}_{\alpha\uparrow}\hat{n}_{\alpha\downarrow},
\end{split}
\end{equation}
where $\hat{c}^{(\dagger)}_{\alpha,\sigma}$ is the fermionic annihilation (creation) on site $\alpha=\mathrm{L,R}$ with spin $\sigma=\uparrow,\downarrow$ and $n_{\mathrm{\alpha}\sigma}=c^\dagger_{\alpha\sigma}c_{\alpha\sigma}$.  
We write $\ket{\sigma,\sigma'} = \hat{c}^\dagger_{\mathrm{L}\sigma}\hat{c}^\dagger_{\mathrm{R}\sigma'}\ket{0}$.
Since $\ket{\uparrow,\uparrow}$ and $\ket{\downarrow,\downarrow}$ both have energy zero and do not couple to any other states, we can write the Hamiltonian in the reduced Hilbert space spanned by $\{\ket{\uparrow\downarrow,0}, \ket{\uparrow,\downarrow}, \ket{\downarrow,\uparrow},\ket{0,\uparrow\downarrow}\}$ as
\begin{equation}
\label{eq:full_doublewell_hamiltonian}
\hat{H}_\mathrm{DW} = \begin{pmatrix}
U + 2\Delta & -t_x & t_x & 0 \\
-t_x & 0 & 0 & -t_x \\
t_x & 0 & 0 & t_x \\
0 & -t_x & t_x & U - 2\Delta
\end{pmatrix}~.
\end{equation}
In the limit $U\gg\Delta,t_x$, an effective low-energy Hamiltonian can be derived with a Schrieffer-Wolff transformation~\cite{duan_controlling_2003}. After projecting out the high-energy double occupancies ($\ket{\uparrow\downarrow,0}$ and $\ket{0,\uparrow\downarrow}$), the Hamiltonian reads
\begin{equation}
\label{eq:effective_hamiltonian_productstate_basis}
\hat{H}_{\text{eff}} = \frac{1}{2}\begin{pmatrix}
-J_\mathrm{ex} & J_\mathrm{ex} \\
J_\mathrm{ex} & -J_\mathrm{ex}
\end{pmatrix},
\end{equation}
where the superexchange energy is given by
\begin{equation}
    J_\mathrm{ex} = \frac{4t_x^2}{U\left(1-(2\Delta/U)^2\right)}.
\end{equation}
This Hamiltonian can also be written in `Heisenberg' form $\hat{H}_\mathrm{eff}=J_\mathrm{ex}\left(\hat{\Vec{S}}_\mathrm{L}\cdot\hat{\Vec{S}}_\mathrm{R} - \frac{1}{4}\right)$, where $\hat{\Vec{S}}_\alpha = \frac{\hbar}{2}\sum_{i,j}\hat{c}^\dagger_{\alpha i}\Vec{\sigma}_{ij}\hat{c}_{\alpha j}$ is the spin operator on site $\alpha$ and $\Vec{\sigma}=(\sigma_x,\sigma_y,\sigma_z)^T$ is the vector of Pauli operators.
The time evolution operator in the Hilbert space spanned by $\{\ket{\uparrow,\uparrow}, \ket{\uparrow,\downarrow}, \ket{\downarrow,\uparrow},\ket{\downarrow,\downarrow}\}$ takes the form
\begin{equation}
    \hat{U}_{\alpha} = \begin{pmatrix}
        1 & 0 & 0 & 0 \\
        0 & (1+e^{i\pi\alpha})/2 & (1-e^{i\pi\alpha})/2 & 0 \\
        0 & (1-e^{i\pi\alpha})/2 & (1+e^{i\pi\alpha})/2 & 0 \\
        0 & 0 & 0 & 1
    \end{pmatrix},
\end{equation}
realising a {\sc $($swap$)^\alpha$} gate, where $\alpha = \frac{1}{\hbar\pi}\int_{\tau_{\text{start}}}^{\tau_{\text{end}}}J_\mathrm{ex}(\tau)\,d\tau$ ~\cite{zhang_scalable_2023}. Note that in our realisation, $t_x(\tau)$ and $\Delta(\tau)$ are periodically modulated (Fig. \ref{fig:S1}b), as is $J_\mathrm{ex}(\tau)$.
In the experiment we are not fully in the limit $U\gg\Delta,t_x$. We therefore use the energy difference of the two lowest eigenstates of the full double-well Hamiltonian (Eq.~\ref{eq:full_doublewell_hamiltonian}) for the calculation of superexchange energy, which is used for the upper $x$-axis in Fig.~\ref{fig:2}g. Fig.~\ref{fig:S1}c shows the resulting $J_\mathrm{ex}$. During the staggered lattice configuration ($\pm 0.25T$), $J_\mathrm{ex}$ is negligible and increases (decreases) by two orders of magnitude as the balanced double-well configuration is entered (exited). The superexchange interaction between atoms located in different double wells is more than four orders of magnitude smaller than that between atoms in the same double well, making it negligible. Finally, Fig.~\ref{fig:S1}d shows the Hubbard $U$, the maximum of $\Delta$ and $t_x$, and the resulting maximum of $J_\mathrm{ex}$ during a pump cycle as a function of lattice depth $V_\text{X}$.

\subsection*{Singlet preparation}
We load an evaporatively cooled, balanced spin mixture of fermionic potassium-40 atoms in the magnetic states $F=9/2$, $m_F=\{-9/2, -7/2\}$ into a crossed dipole trap and further evaporatively cool it, yielding $5.3(2) \times 10^{4}$ atoms at a temperature of 0.102(6) times the Fermi temperature.

\begin{figure}[htbp]
    \includegraphics[width=0.48\textwidth]{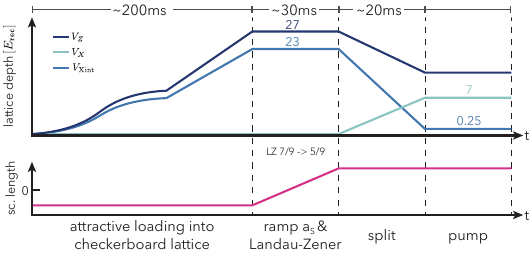}
         \caption{\textbf{Experimental sequence for singlet state preparation}. Loading scheme to prepare strongly interacting atomic singlets on a double well. First, the lattice ($V_{\text{X}},V_{\text{Z}}$) is adjusted to an intermediate strength with a spline-shaped intensity ramp with attractive interactions to maximise the fraction of atoms in doubly occupied unit cells. Then, the lattice powers are increased to their maximum values and held while the scattering length is adjusted to positive values. In the last step, the checkerboard pattern is split into a double-well configuration with an enhanced fraction of atomic singlets by ramps of $V_{\text{X}},V_{\text{Xint}}$ and $V_{\text{Z}}$, respectively. 
         }
    \label{fig:S2}
\end{figure}

The experimental sequence to achieve a high fraction of doubly occupied unit cells with two atoms in the singlet configuration is shown in Fig.~\ref{fig:S2}. First, we use the Feshbach resonance at \SI{201.1}{G} to tune the s-wave interactions between atoms in the $-9/2$ and $-7/2$ sublevels to be strongly attractive. The atoms are then loaded into a shallow chequerboard lattice over 200 ms and subsequently into a deep chequerboard lattice over 10 ms, resulting in a high fraction of paired atoms in the $-9/2$ and $-7/2$ sublevels~\cite{walter_quantization_2023}. To achieve strongly repulsive interactions, we transfer the $-7/2$ population to the $-5/2$ sublevel using a Landau-Zener sweep and adjust the magnetic field to reach the target scattering length. Finally, by ramping down $V_\mathrm{X}$ and ramping up $V_\mathrm{Xint}$, the chequerboard lattice is split, resulting in between 60\% and 75\% of doubly occupied unit cells in the singlet configuration.

\subsection*{Product state preparation}

\begin{figure}[htbp]
    \includegraphics[width=0.48\textwidth]{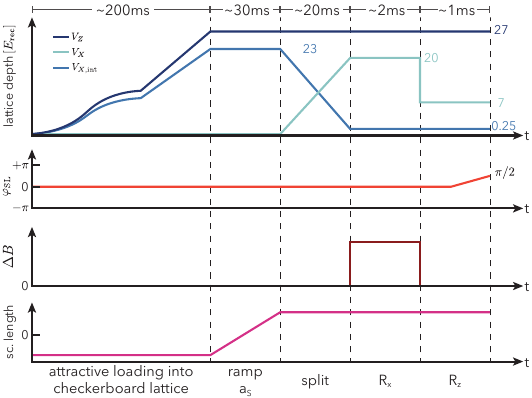}
         \caption{\textbf{Experimental sequence for product state preparation.} 
         To initialise a product state, we follow the conventional singlet preparation procedure until the splitting is performed. Here, in contrast to the singlet preparation, we ramp $V_{\mathrm{X}}$ to a higher value of 20 $E_{\text{rec}}$ and subsequently apply a magnetic field gradient $\Delta B$ to perform a $\pi/2$ rotation $R_x$ around the $x$-axis on the two-particle Bloch sphere, obtaining the $\ket{i_+}$ state [see Fig. \ref{fig:2}(e)]. Then, the lattice depth $V_{\mathrm{X}}$ is lowered and the pump is operated for a quarter pump cycle $T/4$, which in combination performs the rotation $R_z$ on the equator of the Bloch sphere to the desired target state $\ket{\downarrow,\uparrow}$, if the delay between both the pump and the lattice ramp are carefully chosen. 
         }
    \label{fig:S3}
\end{figure}

The loading sequence for the product state $\ket{\downarrow,\uparrow}$ is shown in Fig. \ref{fig:S3}. We start by loading singlets into a deep double-well configuration with the same sequence as in the singlet state preparation, but ramping to different lattice potentials during the split. In this lattice, both tunnellings $t_x$ and $t_x'$ are negligible. Then we apply a magnetic field gradient $\Delta B$, which causes a rotation around the $x$-axis on the two-particle Bloch sphere with singlet state $\ket{s}=\left(\ket{\downarrow,\uparrow}-\ket{\uparrow,\downarrow}\right)/\sqrt{2}$ on the north pole and triplet state $\ket{t}=\left(\ket{\downarrow,\uparrow}+\ket{\uparrow,\downarrow}\right)/\sqrt{2}$ on the south pole [Fig. \ref{fig:2}(d)]. To see this, we write the Hamiltonian in Eq.~\ref{eq:effective_hamiltonian_productstate_basis} in the $\{\ket{s},\ket{t}\}$ basis,
in which it takes the form
\begin{equation}
\hat{H}_{\text{eff}} = \frac{1}{2}\begin{pmatrix}
J_\mathrm{ex} & 0 \\
0 & -J_\mathrm{ex}
\end{pmatrix}.
\end{equation}
The magnetic field gradient couples the $\ket{s}$ and the $\ket{t}$ states (singlet-triplet oscillation, STO),
\begin{equation}
\hat{H}_{\text{STO}} = \frac{1}{2}\begin{pmatrix}
0 & \Delta_{\uparrow\downarrow} \\
\Delta_{\uparrow\downarrow} & 0
\end{pmatrix}~,
\end{equation}
where $\Delta_{\uparrow\downarrow}$ is the energy offset between $\ket{\uparrow,\downarrow}$ and $\ket{\downarrow,\uparrow}$ induced by the gradient.
This can be rewritten as $\hat{H}_\mathrm{STO} =\Delta_{\uparrow\downarrow}\hat{\sigma}_x /2$, where $\hat{\sigma}_{x,z}$ are Pauli matrices. Therefore, a time evolution under this Hamiltonian in the frozen lattice configuration for a time $\tau=\frac{\hbar\pi}{2\Delta_{\uparrow\downarrow}}$ rotates the state vector to the equatorial state $\ket{i_{-}}=\left(\ket{\downarrow,\uparrow}-i\ket{\uparrow,\downarrow}\right)/\sqrt{2}$. We calibrate this duration in a separate measurement. After this rotation, we turn off the magnetic-field gradient, unfreeze the lattice by ramping down $V_{\text{X}}$ and pump for a quarter pump-cycle, such that the integrated $J_\mathrm{ex}$ corresponds to a $\pi/2$ rotation around the $z$-axis of the Bloch sphere, arriving at the target state $\ket{\uparrow,\downarrow}$ in the staggered lattice configuration.

\subsection*{Experimental sequence}

\begin{figure}[htbp]
    \includegraphics[width=0.48\textwidth]{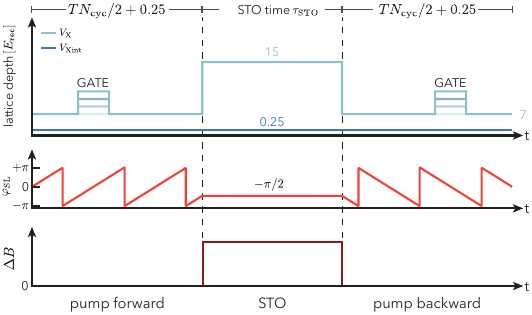}
         \caption{\textbf{Experimental sequence for the STO detection scheme.} After the singlet state preparation, we start the pump by ramping the superlattice phase $\varphi_{\mathrm{SL}}$. We can implement a given gate sequence by changing the lattice depth $V_{\text{X}}$ for the duration of the gate. We proceed to stop the pump in the staggered lattice configuration, increase $V_{\text{X}}$ to freeze the lattice dynamics, and perform STOs by applying a magnetic-field gradient $\Delta B$ for a variable amount of time $\tau_\mathrm{STO}$. After unfreezing the lattice, we reverse the pump for the same number of pump cycles and finally measure the singlet fraction as a function of $\tau_\mathrm{STO}$.}
    \label{fig:S4}
\end{figure}

\begin{figure}[htbp]
\includegraphics[width=0.48\textwidth]{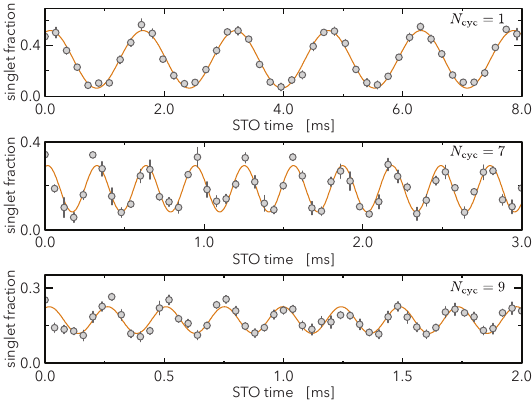}
         \caption{\textbf{STO measurements for different number of operation cycles.} 
         Complementary time traces of the STOs presented in Fig.~\ref{fig:3}(d) of the main text. Here, singlets are pumped apart for different operation cycles $N_{\mathrm{cyc}}$ indicated by the insets. Larger separation ($s=2N_{\mathrm{cyc}}+1$ lattice sites) correspond to faster STOs. The error bars denote the standard error for 3-6 experimental repetitions.}
    \label{fig:S5}
\end{figure}

In Fig. \ref{fig:S4} we show the experimental sequence used for the STO measurements shown in Fig.~\ref{fig:3}, Fig.~\ref{fig:4} and Fig.~\ref{fig:S5}. After state preparation, we start the pump by linearly ramping $\varphi_{\text{SL}}$ with a positive slope $2\pi/T$. We tune the superexchange interaction $J_\mathrm{ex}$ by varying the lattice depth $V_{\text{X}}$ between operation cycles, thereby implementing different gates.
The pump is then halted in the staggered lattice configuration to perform the STO. 
The lattice dynamics is frozen by ramping up $V_{\text{X}}$ and by applying a magnetic-field gradient $\Delta B$ for a certain time $\tau_\mathrm{STO}$. After turning off the magnetic-field gradient and unfreezing the lattice, we reverse the pump by ramping $\varphi_{\text{SL}}$ with a slope $-2\pi/T$ for the same number of pump cycles to bring the atoms back to their initial position. We then measure the nearest-neighbour singlet fraction in the final state as a function of $\tau_\mathrm{STO}$. The last two steps (reversed pump and singlet detection) should be considered together as a detection scheme for long-distance correlation.
The entire sequence can be understood as a many-body atom interferometer. In Fig. \ref{fig:S5} we show complementary time traces for the data shown in Fig. \ref{fig:3} of the main text.  

\subsection*{Detection of singlet fraction and double occupancy}

To measure the fraction of atomic singlets in a doubly occupied unit cell, we first freeze the dynamics by quenching into a deep simple cubic lattice with half the periodicity of the double wells within \SI{100}{\micro\second}. We clean the remaining double occupancies, which are suppressed by the strongly repulsive Hubbard interactions, by transferring all atoms in the -5/2 to the -3/2 state. When another atom in the -9/2 state is present, they will collide and leave the trap. After that, we transfer the remaining -3/2 population back to -7/2. Then, we merge adjacent sites and ramp the Hubbard $U$ to the attractive regime. Singlets form double occupancies in the lowest band, while the Pauli exclusion principle forces triplets to convert to one atom in the lowest band and one in the first excited band.
This enables us to measure the fraction of singlets by detecting the double occupancies in the merged lattice. We sweep the magnetic field over the $-7/2$ and $-9/2$ Feshbach resonance then apply a Landau-Zener RF sweep. The interaction shift then causes only the $-7/2$ population on doubly-occupied sites to be transferred to the $-5/2$ state. The Zeeman sublevels are then separated by applying a magnetic-field gradient and \SI{8}{\milli\second} time of flight \cite{walter_quantization_2023}. 
To measure $\langle\hat{\sigma}_y\rangle$, we apply an additional STO corresponding to a $\pi/2$ and a $3\pi/2$ rotation about the $x$-axis of the two-particle Bloch sphere before merging. This converts the projection on the $y$-axis to the projection on the $z$-axis. We then determine $\langle\hat{\sigma}_y\rangle$ by calculating the difference in singlet fraction for the $\pi/2$ and $3\pi/2$ rotations, normalised by the initial fraction of doubly occupied unit cells.

In Fig.~\ref{fig:2}(c) we directly measure the double occupancies in the simple cubic lattice without cleaning and merging adjacent sites. The double occupancy measurement relies on an energy shift and a spectroscopic transfer into a third hyperfine state, conditioned on the presence of a second atom in the same orbital~\cite{busch_two_1998}.
The double-occupancy detection is thus orbital-selective. To determine the shuttle operation fidelity for single atoms, we calculate the average decay constant, $\beta$, from two exponential fits and convert it to a fidelity using the formula $F=\sqrt{e^{-1/\beta}}$. The square root arises because, in our detection scheme, the failure to shuttle only one atom in a singlet pair is treated as the complete loss of double occupancy, corresponding to two atoms.

\begin{figure}[t!]
\includegraphics[width=0.48\textwidth]{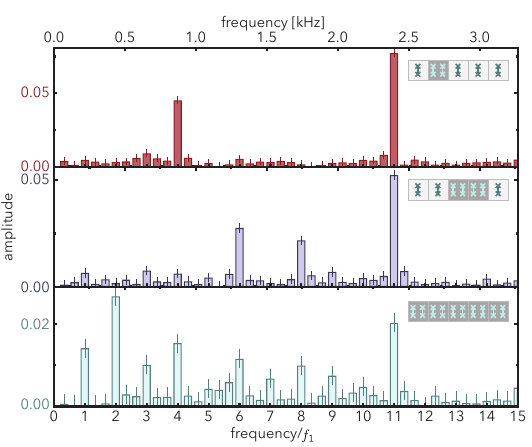}
         \caption{\textbf{Fast Fourier transform of multi-frequency STOs for different gate configurations} The gate sequence used in the measurement can be seen in the top right of each subplot and is identical to the measurements in Fig.~\ref{fig:4}(a-c). 
         The lower x-axis is given in frequency units normalised to the base frequency $f_1=\SI{218(1)}{\hertz}$, which comes from a separate calibration measurement. The upper $x$-axis shows the bare frequency. The dominant frequency peaks appear at integer values of the base frequency due to the quantised displacement in the lattice. The error bars represent the FFT amplitude uncertainty, calculated by propagating the standard error of the time-domain signal through the discrete Fourier transform sum.
         }
    \label{fig:S6}
\end{figure}

\subsection*{Analysis of STO with multi-frequency fit and FFT}

Since the energy difference under a homogeneous magnetic gradient between $\ket{\uparrow,\downarrow}$ and $\ket{\downarrow,\uparrow}$ states is linear in the separation of the atom pair, we expect the time-trace of the STO to be a superposition of sine-waves with frequencies $sf_1$, where $s\in\mathbb{N}^{+}$ is the atom pair separation in number of lattice sites and $f_1$ is the base STO frequency at a distance of one lattice site. We use the fit function \begin{equation}
\begin{split}
    F_\mathrm{singlet}(\tau) =~ & e^{-\Gamma\tau}\sum_{s=1}^{12} A_s\sin(2\pi sf_1\tau+\theta_s)\\ 
    &+ F_0.
\end{split}
\end{equation}
The fit parameters are $\Gamma,A_s,\theta_s$ and $F_0$. We also calculate the Fast Fourier transform (FFT) of the time traces in Fig.~\ref{fig:4} to validate our fit function. The FFT for different gate sequences can be seen in Fig.~\ref{fig:S6}. The dominant peaks are all at integer multiples of the base frequency $f_1$, which justifies the choice of fit function.

\subsection*{Comparison with tweezer transport}
To place our topological-pumping platform in context, we discuss three key features compared to state-of-the-art neutral-atom tweezer arrays.

\textit{High qubit density \& native swap capability.} Dense qubit packing is essential when the field of view and numerical aperture constrain the usable area for large-scale atomic arrays. Optical lattices naturally support sub-micrometre site spacing (\SI{532}{nm} in this work), resulting in qubit densities nearly two orders of magnitude higher than typical tweezer arrays with a few micrometres spacing~\cite{browaeys_many-body_2020}. However, routing atoms in such dense arrays poses a challenge as the space between sites is insufficient to fit additional tweezer beams without crosstalk. In contrast, our two-band topological pump inherently swaps qubits by routing them through orthogonal orbitals, effectively letting atoms `pass through' one another, making it ideal for densely filled arrays. 

\textit{Robust ground-state-preserving transport.} While tweezer arrays can achieve high transport fidelities in free-space motion, their performance degrades significantly when atoms are moved through an underlying lattice potential for dense qubit storage~\cite{gyger_continuous_2024}. In contrast, our topological pump is designed to operate natively within a dense lattice array, maintaining high fidelity (single shift $>99.7\%$) and thereby eliminating the typical trade-off between qubit density and transport reliability. Crucially, the transport also preserves motional coherence, as atoms remain in the motional ground state throughout the pump. Conversely, tweezer-based shuttling can leave atoms in a motionally excited state of the tweezer or underlying lattice after transport.

\textit{Competitive transport speed.}  
The operation time for a single shift gate in this work is \SI{1.25}{ms}. This timescale is currently limited by the bandwidth of our phase lock, which is purely technical. The fundamental physical limit to the gate time, however, is set by the requirement for adiabaticity, which dictates that the evolution time must exceed the inverse of the energy gap. In deeper lattices, where the energy gap can be increased to tens of kilohertz, we anticipate reducing the gate time significantly, potentially reaching $\SI{100}{\micro\second}$ and less. For comparison, state-of-the-art tweezer-array platforms report transport speeds of 0.55\,$\mu$m/$\mu$s~\cite{bluvstein_quantum_2022}, corresponding to tens of microseconds for moving an atom to a nearby one, or hundreds of microseconds to shuttle atoms between different functional zones for storage, entangling, and readout~\cite{bluvstein_logical_2023}. However, we emphasize that such speeds are typically achieved via free-space transport. In scenarios involving a background lattice for dense qubit storage, the transport speed drops by up to an order of magnitude~\cite{gyger_continuous_2024}, in order to mitigate motional heating.

\subsection*{Error analysis}

\textit{Single-particle transport errors.} In our topological pump, the principal single-particle errors that limit the pump efficiency arise from undesired residual tunnelling (dispersive loss). 
For the lattice depths used, the inter-dimer tunnelling in the dimerised configuration is below \SI{10}{Hz}. In the staggered configuration, tunnelling between neighboring sites (approximately \SI{180}{Hz}) is suppressed by a large energy site offset (approximately \SI{4.2}{kHz}). Throughout the pump cycle, the tunnellings along transverse directions and the next-nearest-neighbor (NNN) tunnellings are below \SI{10}{Hz}. More importantly, we utilize an external harmonic potential with a trapping frequency of about \SI{90}{Hz}.
The local tilt introduced by the trap acts as Wannier-Stark localization. This tilt increases linearly with the distance from the trap center, exceeding the relevant tunnellings as close as two lattice sites from the center and thus significantly suppresses the tunnelling process across the entire cloud~\cite{zhu_reversal_2024}.
Any spread of the wavefunction due to residual tunnellings would manifest as loss of double‐occupancy during pumping, and our measurement in Fig.~\ref{fig:2}(c) bounds this dispersive loss to $\lesssim0.2\%$ per pump cycle. 
Band‐excitations and motional heating from non‐adiabaticity are also negligible, since the pump frequency ($1/\text{cycle time}=\SI{400}{Hz}$) is chosen to be much smaller than the band gap. Using the Landau–Zener formula, we estimate the excitation probability \(P_e\) when the energy gap reaches its minimum during the pump cycle:
\begin{equation}
P_e = \exp\!\left(-\pi^2 \,\frac{E_{\mathrm{gap}}^2}{\partial\nu/\partial\tau}\right).
\end{equation}
In our typical lattice configuration, the minimum gap occurs in the dimerized configuration, where $E_{\mathrm{gap}} = 2t_x \approx \SI{2.6}{kHz}$. The sweep rate of the energy offset is $\partial\nu/\partial\tau = 2\,\partial\Delta/\partial\tau \approx \SI{9}{kHz/ms}$. This yields an excitation probability of $P_e \approx 6 \times 10^{-4}$.
Together, these effects result in a single-step pump fidelity of $99.78(3)\%$. There are several straightforward technical improvements which will enable efficiencies on the order of $\geq99.9\%$. First, one can implement a true linear (Wannier–Stark) potential for uniform tilt to suppress tunnelling throughout the system.
Second, deeper lattices can be used to further suppress all undesired tunnellings to sub-Hertz regime, while keeping the relevant intra-dimer tunnelling processes the same.
Third, deeper lattices can also be used to enlarge the band gaps and improve the degree of adiabaticity.

\textit{Two-qubit gate errors.} The dominant technical errors in our two‐qubit superexchange gates stem from fluctuations in lattice depth, inhomogeneity from beam‐profile, and magnetic‐field jitter, all of which affect $J_{\mathrm{ex}}\propto t^2/U$.
We measure low‐frequency ($<1/T_\mathrm{gate}$) noise levels (RMS) of 1\% in $V_X$, 3\% in $V_{X_\mathrm{int}}$, and 0.3\% in $V_Z$, plus about $0.6\%$ variation across the cloud due to the Gaussian beam profile, and $1\%$ fluctuation (RMS) in Hubbard $U$ from magnetic‐field instability (TABLE~\ref{table:gate_error}).
Propagating these gives a total $J_{\mathrm{ex}}$ uncertainty of approximately $7\%$. This value is consistent with the gate fidelity of $\simeq93.6\%$ estimated from the decay constant ($\approx15$ operation cycles) of singlet–triplet oscillations in Fig.~\ref{fig:3} and Fig.~\ref{fig:S5}, assuming that our lattice sites are, on average, half-populated such that each singlet pair is involved in roughly one two-qubit gate per operation cycle. The total circuit depth is primarily limited by the fidelity of two-qubit gates, while the atomic lifetime limited by background gas collisions is on the order of one minute, which is not a limiting factor of the achievable circuit depth. These technical errors for the two-qubit gate are not fundamental—improvements such as laser‐power stabilization, magnetic‐field feed‐forward control, can readily push two‐qubit gate fidelities beyond 99.3\%~\cite{yang_cooling_2020}.
Furthermore, there exist established schemes for realising {\sc $($swap$)^\alpha$} gates that are based on mechanisms inherently insensitive to fluctuations in tunnelling parameters~\cite{anderlini_controlled_2007,kaufman_entangling_2015}. Implementing such approaches could significantly improve the performance and robustness of our quantum circuit.

{\renewcommand{\arraystretch}{1.2}
\begin{table}[h]
  \centering
  \begin{tabular}{lcc}
    \hline\hline
    Error source & Level[\%] & Impact on $J_{\mathrm{ex}}[\%]$ \\
    \hline
    Lattice‐depth noise in $V_\mathrm{X}$           & 1     & 4    \\
    Lattice‐depth noise in $V_{\mathrm{Xint}}$       & 3     & 5    \\
    Lattice‐depth noise in $V_\mathrm{Z}$                      & 0.3   & 0.5  \\
    Lattice inhomogeneity           & 0.6   & $1$  \\
    Fluctuations in Hubbard $U$               & 1     & $1$    \\
    \hline
    {\bf Total propagated error}                      &         & {\bf $\approx7$} \\
    \hline\hline
  \end{tabular}
  \caption{Technical error sources and their impact on the superexchange gate fidelity.}
  \label{table:gate_error}
\end{table}
}
\end{document}